\documentclass[pdflatex,sn-mathphys-num]{sn-jnl}
\usepackage{graphicx}%
\usepackage{multirow}%
\usepackage{amsmath,amssymb,amsfonts}%
\usepackage{amsthm}%
\usepackage{mathrsfs}%
\usepackage[title]{appendix}%
\usepackage{xcolor}%
\usepackage{textcomp}%
\usepackage{manyfoot}%
\usepackage{booktabs}%
\usepackage{algorithm}%
\usepackage{algorithmicx}%
\usepackage{algpseudocode}%
\usepackage{listings}%
\usepackage{bm}  
\theoremstyle{thmstyleone}%

\theoremstyle{thmstyletwo}%

\theoremstyle{thmstylethree}%

\begin{document}

\title[Article Title]{3D Stokes polarimetric imaging at nanoscales.}

\author[1]{\fnm{Isael} \sur{Herrera}}\email{isael.herrera@fresnel.fr}

\author*[1,2]{\fnm{Miguel A.} \sur{Alonso}}\email{miguel.alonso@fresnel.fr}

\author*[1]{\fnm{Sophie} \sur{Brasselet}}\email{sophie.brasselet@fresnel.fr}

\affil[1]{\orgname{ Aix Marseille Univ, CNRS, Centrale Med},\orgdiv{Institut Fresnel},  \orgaddress{ \city{Marseille}, \postcode{13013}, \country{France}}}

\affil[2]{\orgdiv{The Institute of Optics}, \orgname{University of Rochester}, \orgaddress{\city{Rochester}, \postcode{14627}, \state{NY}, \country{USA}}}

\abstract{ 

Optical fields polarized along three dimensions are frequent in optical microscopy and nanophotonics, and yet 
retrieving their polarization distribution is challenging.
We present the experimental implementation of three-dimensional (3D) Stokes polarimetric imaging of nonparaxial optical fields with nanoscale spatial resolution. This approach extends classical Stokes polarimetry --traditionally limited to paraxial fields-- into the nonparaxial regime. We use an array of gold nanospheres, each acting as a localized electric dipolar scatterer, to probe 3D polarization states over a field of view of tens of micrometers. The scattered signal is collected by a high numerical aperture objective lens and separated into its circular polarization components, providing a very simple imaging system.
We introduce a computational algorithm to efficiently extract the physical parameters from the generated dipole spread functions with a high throughput across the whole field of view. Finally, we show that this method can also be applied 
to single-molecule localization and orientation fluorescence microscopy.}

\keywords{Non-paraxial polarimetry, Stokes-Gell-Mann parameters, Polarization microscopy, single nanosphere scatterer}



\maketitle

\section{Introduction}\label{sec1}
Polarimetry is a well-established technique for characterizing the vectorial oscillations of paraxial light, with widespread applications including in biomedical imaging \cite{He2021}, quantum optics \cite{Goldberg2022}, and astronomy %
\cite{Akiyama_2021}. 
It relies on sequential \cite{Schaefer2007} or single-shot \cite{Zaidi2024, Pierangeli_2023, Ramkhalawon:13} intensity measurements to estimate the four Stokes parameters that define transverse polarization states
\cite{Alonso2023}.
There is, however, growing interest in characterizing polarization beyond the paraxial regime where the field can oscillate in all three dimensions, e.g. under strong focusing  or in confined geometries near sources, interfaces or nanostructures. In this regime, fully polarized light corresponds to the electric field tracing periodically an ellipse in 3D with a given handedness, while partial polarization implies more complicated field oscillations \cite{Alonso2023}.  
The nonparaxial generalization of the Stokes formalism has been the subject of many studies \cite{Soleillet1929,Gil2004, Ellis2005, Dennis2007, Sheppard2012, Bjrk2014, Alonso2023}.
It 
requires nine analogues of the Stokes 
parameters, referred to here as the Stokes-Gell-Mann (SGM) parameters \cite{Alonso2023}. 

Measuring nonparaxial polarization
is challenging because the fields contain longitudinal components and may vary within nanometric scales. Experimental methods include far-field measurements that allow inferring the near-field polarization distribution by assuming full coherence and polarization \cite{herrera2023measurement,quinto2023interferometric,kumar2024experimental,maluenda2025characterization}. Local measurements have been proposed 
based on nanoprobes, such as near-field tips in the linear \cite{Grosjean2010,Kabakova2016,Neugebauer2018,leFeber2014,leFeber2019,Rotenberg2014} or nonlinear \cite{Frischwasser2021} optical regimes, functionalized tips with a scattering nanoparticle \cite{Lee2007,Ahn:09}, or a single isolated metallic nanoparticle positioned at a coverslip's surface \cite{Bauer:2014,Bauer:2015,Neugebauer2018,Yang2023}. Indirect approaches have been based on diffraction by metallic edges \cite{Marchenko2011,Huber2013}, material modifications at the nanoscale \cite{Grosjean:06,Porfirev2022}, fluorescent layers \cite{Otte:2019} and single molecules \cite{Sick2000,novotny2001,Lieb2004}, or electrical signals using p-n junctions of nanometric size \cite{Yu2022}.  
The 
use of scattering by a nanoparticle is particularly interesting, since it allows direct retrieval at the nanoparticle's position \cite{Bauer:2014,Bauer:2015,Eismann2020} of the complex local electric and magnetic field vectors when using transmission scanning interferometry \cite{Neugebauer2018,elec_mag}. Its implementation, however, has so far been based on measuring the radiation pattern at the objective lens' back-focal plane, hence spreading the information over many pixels and limiting the analysis to single nanoparticles in focused fields.

In this work, we present a single-shot nonparaxial polarimetric imaging modality.
We employ a 2D array of gold nanospheres to probe the field's local polarization at many points across a field of view (FOV) of 
tens of micrometers. 
The nanospheres' subwavelength size 
causes their induced dipoles to be approximately proportional to the local external electric field.  
Their radiation is imaged in the dark field mode at high numerical aperture, and projected onto its left- and right-hand circular (LHC and RHC, respectively) polarization components, forming for each particle a so-called dipole spread function (DSF) pair. We refer to this 3D Stokes polarimetric imaging approach as \textit{circular-projection 3D polarimetry} (C3Pol). 
We also
exploit the linear dependence of the measured DSFs on the SGM parameters to retrieve these using a fast, direct computational procedure. 
We demonstrate C3Pol's performance theoretically and experimentally on spatial distributions of complex optical fields with different 3D polarizations.
Finally, we apply C3Pol in fluorescence imaging to measure 
the 3D orientation and localization of single molecules, with  performance comparable to that of the most efficient single molecule orientation and localization microscopy (SMOLM) methods \cite{Brasselet:23}.

\section{Results}\label{sec2}
\subsection{Complete nonparaxial polarimetry via circular polarization projection.}

The polarization matrix $\mathbf{\Gamma}$ for a nonparaxial electric field $\mathbf{E}$ is defined as the correlation matrix with elements ${\Gamma}_{i,j}=
\langle\mathrm{E}_{i}\mathrm{E}_{j}^{*} \rangle$, where $\langle\cdot\rangle$ denotes time averaging over the detector's integration time. This matrix can be parameterized in terms of the SGM parameters \cite{Alonso2023,Herrera2024} (Supplementary Note 1):
\begin{equation}
  \begin{aligned}
   & \quad  \mathbf{\Gamma}
 =\frac{1}{2}\left(\begin{array}{ccc}\frac{2}{3}S_0 +S_{11}+\frac{S_{12}}{\sqrt{3}}& S_{23}-iS_{33} & S_{22}+iS_{32}   \\S_{23}+iS_{33} & \frac{2}{3}S_0 -S_{11}+\frac{S_{12}}{\sqrt{3}}&  S_{21}-iS_{31}   \\ S_{22}-iS_{32} & S_{21}+iS_{31}   &  \frac{2}{3}S_0 -2\frac{S_{12}}{\sqrt{3}}
\end{array}\right).
       \label{eq:Int_S3}
 \end{aligned}
\end{equation}

\begin{figure}[h]
    \centering
    \includegraphics[scale=0.35]{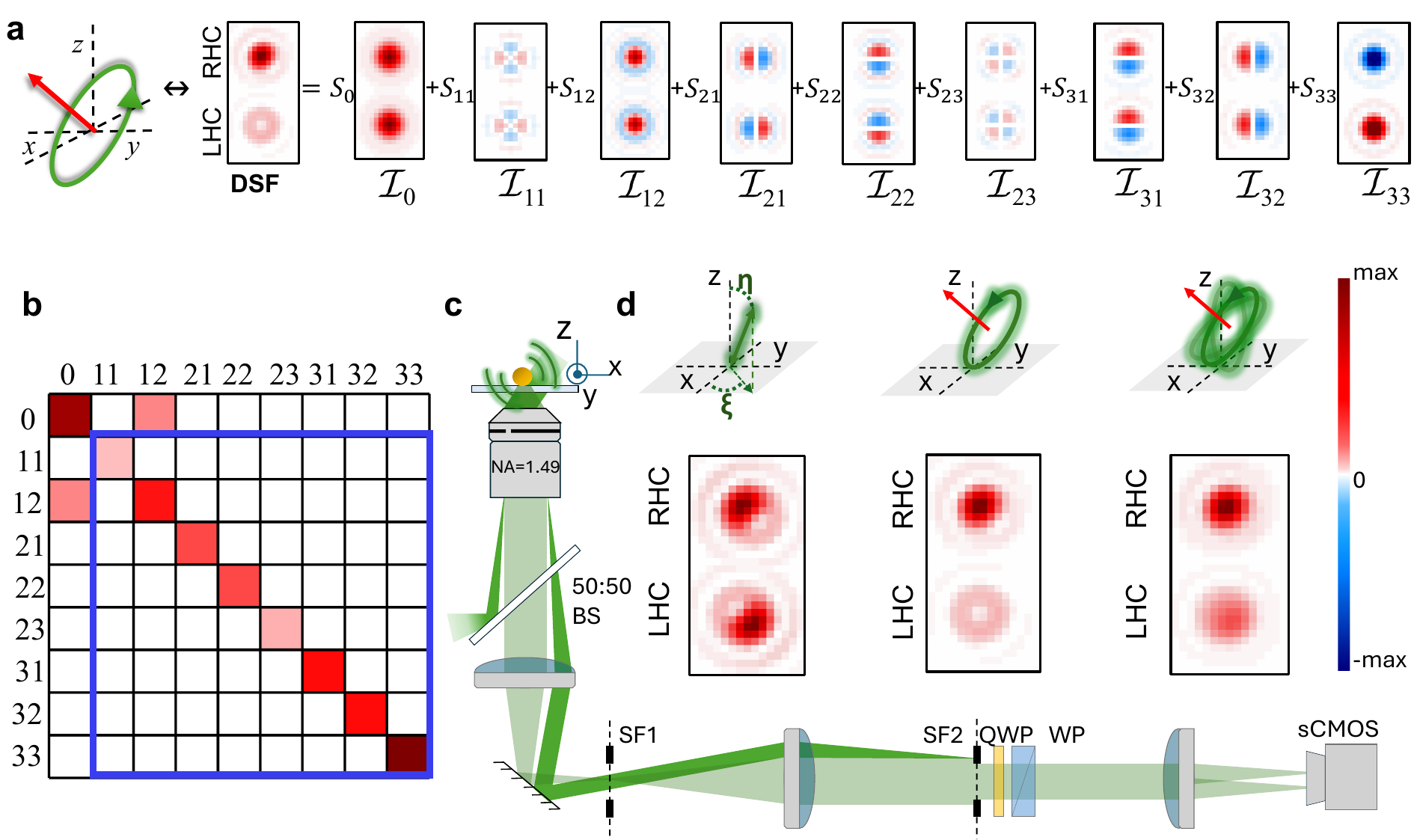}
    \caption{\textbf{3D Stokes polarimetric imaging principle}. \textbf{a} Illustration of a fully polarized elliptical dipole ($P=1$) (red arrow: spin vector), whose major axis has a polar angle $\eta=45^{\circ}$ and an in-plane angle $\xi=45^{\circ}$, and whose minor axis is along $x$ with aspect ratio $r_{ba}=0.5$. Decomposition of the corresponding Circ-DSF on the SGM basis $\{\mathcal{I}_{j}\}$. \textbf{b} Matrix containing the inner product of the basis elements weighted by the inverse of $\mathcal{I}_0$, calculated for the fully unpolarized state ($S_{j \neq 0}=0$). The sub-matrix enclosed in blue is a representative case of the FIM. \textbf{c} Schematic of the set-up. BS: unpolarized (50:50) beam splitter; SF1,SF2: spatial filters; QWP:  quarter-wave plate; WP: Wollaston prism. \textbf{d} Illustration of three different polarization states  and their corresponding Circ-DSFs. From left to right: a fully polarized ($P=1$) rectilinear dipole with angles $\eta=45^{\circ}$, $\xi=45^\circ$; a fully ($P=1$) and a partially ($P=0.6$) polarized dipoles whose main eigenvector's major axis is oriented at $\eta=45^{\circ}$, $\xi=45^{\circ}$, its minor axis is parallel to the $x$-$y$ plane, and with aspect ratio of $r_{ba}=0.5$. The basis, FIM and DSFs were computed assuming in-focus dipoles in a continuous medium with refractive index $n_0=1.518$, an objective aperture of $\mathrm{NA}=1.49$, a wavelength $\lambda=532$ nm and a pixel size of 52.72 nm. }   
    \label{fig:Fig1}
\end{figure}

Here, $S_0$ is proportional to the local intensity, $S_{1i}$ $(i=1,2)$ and $S_{2i}$ $(i=1,2,3)$ describe the 3D shape and orientation of the region explored by the field,
and the vector $(S_{31},S_{32},S_{33})$ is proportional to
the \textit{spin vector}. For fully polarized fields, the field traces an ellipse whose aspect ratio is denoted as $r_{ba}$, and the spin vector is perpendicular to the ellipse's plane (Fig. \ref{fig:Fig1}a).

In C3Pol, the polarization information 
is encoded in the field scattered by gold nanospheres that act as Rayleigh scatterers
(Supplementary Fig. S1). In this regime, the induced dipole $\bm{\mu}$ 
is approximately proportional to the external electric field, $\bm{\mu} \propto \textbf{E}$, so that $\mathbf{\Gamma}=\langle\bm{E}\bm{E}^{\dagger} \rangle$ is approximately proportional to the dipole second moment matrix $\mathbf{m}=\langle\bm{\mu}\bm{\mu}^{\dagger} \rangle$, where $\dagger$ denotes complex conjugation and transposition. 
Polarization is measured simultaneously at several locations with nanometric precision from each of the scatterers' DSFs 
(Supplementary Note 1 and Fig. S2), each expressible as a linear superposition 
of a set of components $\mathcal{I}_j$ (referred to as the SGM basis) weighted by the corresponding SGM parameter $S_j$ \cite{Herrera2024,Curcio2020} (Supplementary Note 1):
\begin{equation}
    \mathbf{DSF} = \sum_j S_j \mathcal{I}_j, \quad j \in \{0,11,12,21,22,23,31,32,33\}.
\label{eq:Eq2}
\end{equation}
This decomposition, which provides a simple estimator for the SGM parameters, is general and valid for any microscope, the explicit form of the SGM  basis functions $\{ \mathcal{I}_j\}$ being determined by the microscope configuration (Fig. \ref{fig:Fig1}a). Intuitively, a microscope capable of complete nonparaxial polarimetry is one whose functions $\{ \mathcal{I}_j\}$ are linearly independent and of comparable magnitude. 

We now show that splitting each DSF into two circular polarization components 
suffices to encode in their shape the complete 3D polarization information. 
Figure \ref{fig:Fig1}a illustrates the RHC and LHC 
components of the SGM basis, referred to as Circ-DSFs, which are linearly independent under the inner product $\langle \mathcal{I}_j \mathcal{I}_k\rangle_{{\rm pix},p}$, where $\langle \cdot \rangle_{{\rm pix},p}$ indicates summation over all detector pixels of both polarization channels.
Figure~\ref{fig:Fig1}b shows the matrix $\langle \mathcal{I}_j \mathcal{I}_k / \mathcal{I}_0 \rangle_{{\rm pix},p}$  
which is a representative case of the Fisher Information Matrix (FIM) that quantifies the DSFs' polarization information content  \cite{Herrera2024}. This matrix is almost diagonal, showing that there is nearly no correlation in the estimation of the SGM parameters (except between $S_0$ and $S_{12}$) when using circular polarization projection. In contrast, when using no polarization projection or linear polarization projection
$\mathcal{I}_{21} = \mathcal{I}_{22} = \mathcal{I}_{33} = 0$ 
(Supplementary Fig. S3),
so the resulting DSFs are independent of $S_{21}$, $S_{22}$ and $S_{33}$, and hence 
do not lead to full
3D polarimetry.

In the C3Pol Stokes polarimetric imaging 
setup (Fig. \ref{fig:Fig1}c), a high numerical aperture objective ($\mathrm{NA}=1.49$) 
generates 
the test field and collects the scattered light.
The total magnification of 125$\times$ provides sufficient pixelization of the Circ-DSFs imaged separately on a camera 
(Methods). 
Figure \ref{fig:Fig1}d shows examples of simulated 
Circ-DSFs 
for a fully polarized rectilinear dipole, and for fully and partially polarized dipoles with nonzero spin (calculation details in Supplementary Note 1; parametrization description in Supplementary Note 2, Fig. S2; more examples in  
Supplementary Figs. S4-S6). For rectilinear dipoles (Fig. \ref{fig:Fig1}d, left), both Circ-DSFs contain equal power, their shape depends on the polar angle $\eta$, and their orientation depends on the in-plane angle $\xi$. For dipoles with spin (Fig. \ref{fig:Fig1}d, middle), the axial spin component ($\propto S_{33}$) causes a power and shape imbalance between the channels.

\subsection{ Stokes-Gell-Mann formalism for parameter estimation.}

\begin{figure}[h]
    \centering
    \includegraphics[scale=0.35]{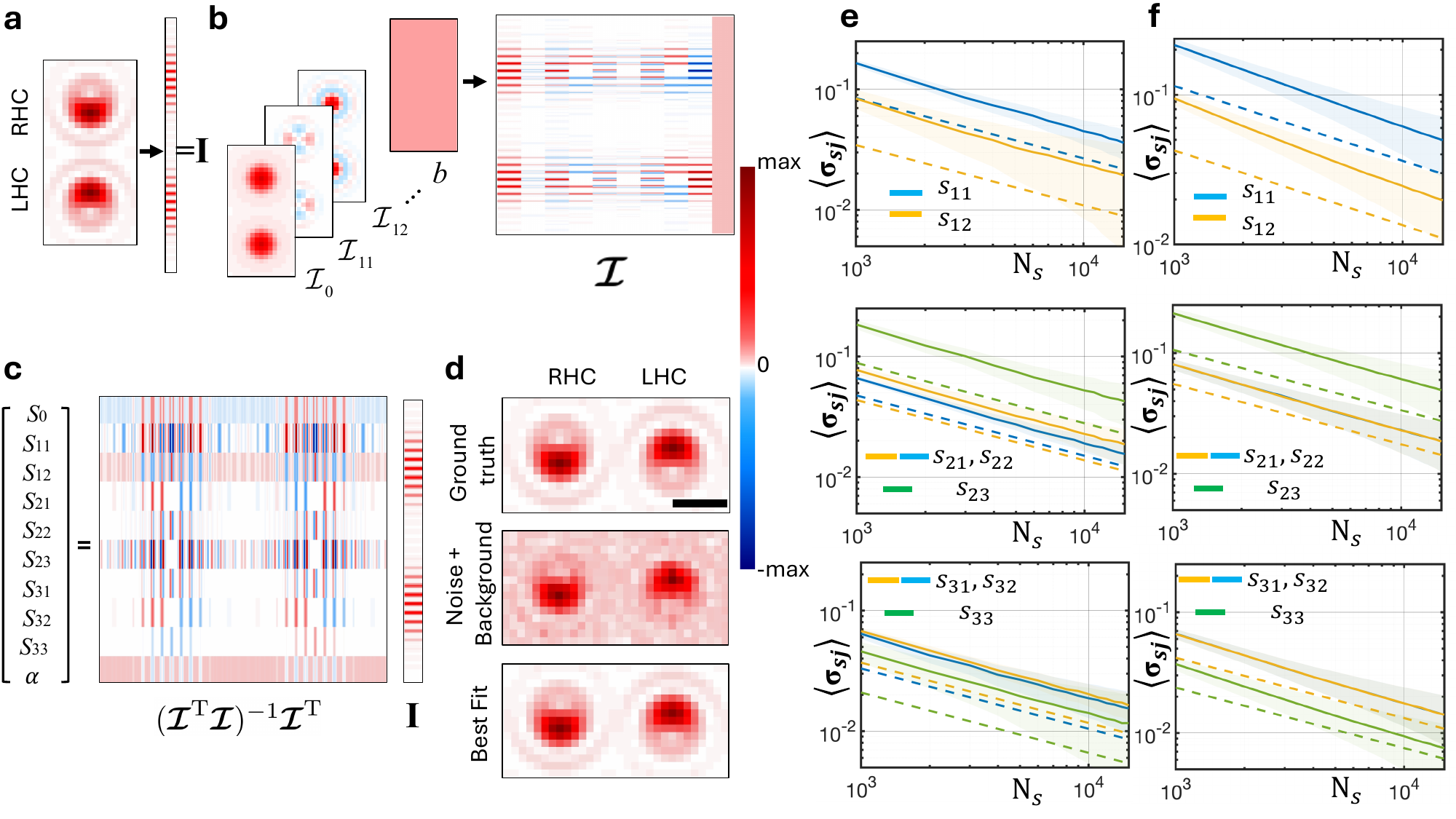}
    \caption{ \textbf{SGM parameter estimation illustration and precision limits. a} Vector form of the combined Circ-DSFs. \textbf{b} Construction of the matrix $\mathcal{I}$, including a constant column accounting for a uniform background. \textbf{c} Illustration of the method for SGM parameter retrieval. \textbf{d} Top: synthetic ground-truth  Circ-DSF pair; Middle: synthetic Circ-DSF pair containing 7500 signal photons and a uniform background noise of 10 photons per pixel; Bottom: best fit using the retrieval procedure described in the text and illustrated in \textbf{c}. \textbf{e} and \textbf{f} Estimation precision for the normalized SGM parameters in terms of the number of signal photons, for fully and partially polarized states, respectively. The dotted lines show the CRLB, while the solid lines show the mean precision obtained with Monte Carlo simulations and the shaded areas indicate the corresponding standard deviation. Due to the system's axial symmetry, the graphs for each pair $(s_{21}, s_{22})$ and $(s_{31}, s_{32})$ overlap. The Circ-DSFs and the basis were computed assuming in-focus dipoles in a continuous medium with refractive index $n=1.518$, an effective aperture at the detection path of $\mathrm{NA}=1.49$, $\lambda=532$ nm and a pixel size of 52.72 nm. }   
    \label{fig:Fig2}
\end{figure}

To arrive at a computationally efficient parameter retrieval algorithm,
we initially assume that the emitter's position $(x_0,y_0)$ matches the center of the selected image sections containing each Circ-DSF component. These sections are merged into a single 1D array (Fig. \ref{fig:Fig2}a), which can be written as
$\mathbf{DSF}=\bm{\mathcal{I}}\cdot  {\boldsymbol{S}}$, where $\bm{\mathcal{I}}$ is a rectangular matrix whose columns are 
the 1D array forms of the SGM basis elements supplemented by a constant column accounting for a uniform background (Fig. \ref{fig:Fig2}b), while $\boldsymbol{S}$ is a vector containing the SGM parameters and a constant $\alpha$ accounting for the background contribution. Due to the linear independence of the columns of $\bm{\mathcal{I}}$, we can calculate 
the pseudo-inverse $\bm{\mathcal{I}}^{-1}=(\bm{\mathcal{I}}^{\rm T}\bm{\mathcal{I}})^{-1}\bm{\mathcal{I}}^{\rm T}$,   T denoting transposition. The SGM parameters are then estimated as $\boldsymbol{S}=\bm{\mathcal{I}}^{-1}\cdot\mathbf{DSF}$, from which we find $\mathbf{\Gamma}$ (equation \ref{eq:Int_S3}), as well as its eigenvalues $\Lambda_j$ and eigenvectors $ \mathbf{e}_j$ (Supplementary Note 2 and Methods). 
For fully polarized fields, the dipole is approximated as the eigenvector corresponding to the largest eigenvalue, $\bm{\mu}\approx \mathbf{e}_1$, (Methods).
Otherwise, the retrieved matrix $\mathbf{\Gamma}$ can be transformed into that of a physically realizable polarization state by replacing any retrieved negative eigenvalue with zero (Methods). Figure \ref{fig:Fig2}d illustrates that this retrieval algorithm leads to the expected results even in the presence of noise and background.

The expected SGM parameter retrieval precision is quantified by the CRLB for the normalized SGM parameters $s_j=\frac{\sqrt{3} S_j}{2 S_0}\in [-1, 1]$ 
 (Supplementary Note 3) \cite{Herrera2024}. The CRLB mean values sampled over $s_j$ are shown in 
Figs.~\ref{fig:Fig2}e and \ref{fig:Fig2}f for fully and partially polarized states, respectively. These values display the highest precisions for the $s_{33}$ component, which also exhibits the highest value in the FIM (Fig. \ref{fig:Fig1}b), and a relatively more intense element $\mathcal{I}_{33}$ (Fig. \ref{fig:Fig1}a). 
The lowest precision results for the estimation of the in-plane components $s_{11}$ and $s_{23}$, as also visible in the FIM and corresponding SGM basis components. C3Pol is therefore less sensitive to in-plane dipole orientation,
while off-plane orientation and spin can be estimated with high sensitivity. Note that, beyond $\sim1000$ signal photons, the precision is an order of magnitude smaller than the largest possible value for $s_j$, and that the estimation of the normalized SGM parameters is more precise for fully polarized states (Fig. \ref{fig:Fig2}e,f).

The estimator 
$\bm{\mathcal{I}}^{-1}$ (a simple least-squares fit) is unbiased and reaches the CRLB when noise is Gaussian; however, noise in optical measurements is Poissonian. Previous work \cite{Goudail2017, Vio2005} argues for 
the convenience of least squares
against maximum likelihood estimation, best suited for Poisson noise but that can be computationally expensive without offering significant improvements for large numbers of photons.  
To quantify the precision and accuracy of the estimator $\bm{\mathcal{I}}^{-1}$, we performed Monte Carlo studies in conditions like those assumed in the examples in Fig.~\ref{fig:Fig1} (Supplementary Note 4).
The results, presented as solid lines in Fig. \ref{fig:Fig2}e-f, follow the CRLB behavior with slightly higher values. Simulations show high accuracy in SGM parameter estimation (Supplementary Figs. S7,S8), with a bias for all parameters typically below $0.05$ for 5000 signal photons (and below 0.02 for fully polarized states).
The accuracy and precision for the orientation angles, ellipse aspect ratio $r_{ba}$ and degree of polarization $P$ follow from those for the SGM parameters. 
For rectilinear dipoles (Supplementary Figs. S9,S10), the estimation of orientation angles reaches both precision and accuracy below $\sim 2^\circ$, while the estimation of $P$ is better than $\sim 0.05$ in precision and bias. For elliptical dipoles, similar performance is obtained for orientation and $P$, while $r_{ba}$ reaches precision and bias below $\sim 0.05$ (Supplementary Figs. S11-S14). 

The Monte Carlo study just discussed assumes in-focus dipoles centered exactly at the Circ-DSFs' central pixel, which is highly unlikely in practice. 
The relative position of the two Circ-DSFs can also be affected by the registration process at each polarization channel (Supplementary Note 5 and Fig. S15). To account for this possible shift in the retrieval procedure, we developed an algorithm to estimate the emitter's position in both polarization channels, based on using different bases associated with different 3D positions (Methods), 
This method allows 3D localization estimation for single emitters with precisions of 13 nm laterally and 75 nm longitudinally (Methods).

\begin{figure}[h]
    \centering
    \includegraphics[scale=0.145]{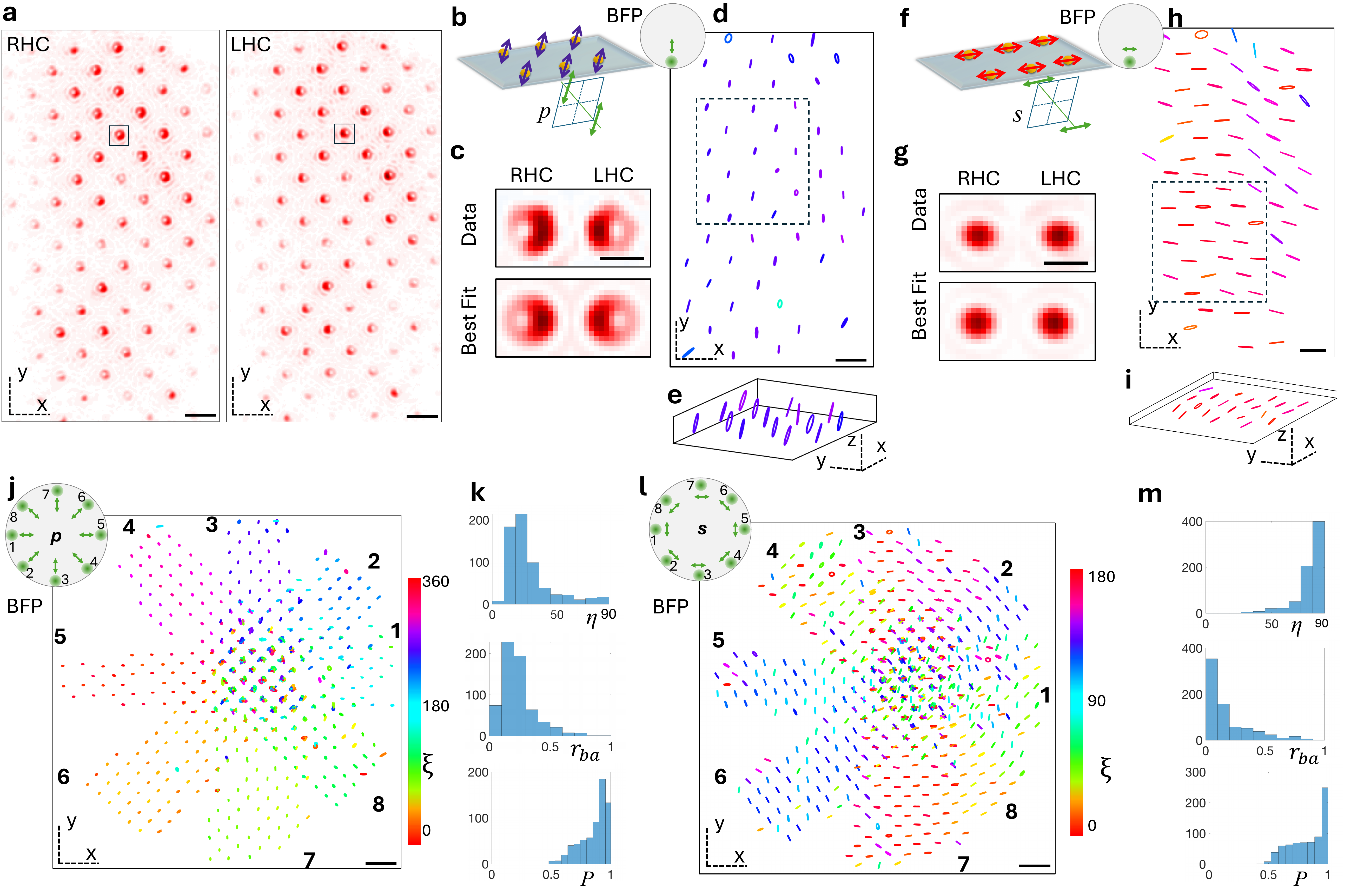}
    \caption{\textbf{C3Pol on 3D linearly polarized states.} \textbf{a} C3Pol images of the nanosphere array. Scale bar: 1.5 $\mu$m. \textbf{b} and \textbf{f} Illustration of the generation of uniform p- and s-polarized fields (inset: BFP scheme). \textbf{c} and \textbf{g}, representative examples of measured Circ-DSFs and their best fit for incident p- and s-polarized fields, respectively. Scale bar: 365 nm. \textbf{d} and \textbf{h} Top view of 2D localized elliptical dipoles generated by p- and s-polarized fields; the ellipse's color indicates the value of the major axis' in-plane angle $\xi$. Scale bar: 1.5 $\mu$m. \textbf{e} and \textbf{i} Lateral view of the measured 3D oriented polarization ellipses within an area of $5.8 \mu$m$\times6.3\mu$m located at the central part of the images shown in \textbf{d} and \textbf{h}, respectively.  \textbf{j} and \textbf{l}, image composed of the detected ellipses generated with eight incident angles (see insets), for p- and s-polarizations, respectively. In \textbf{j} the in-plane angle $\xi$ is color-coded from $0$ to $360^{\circ}$, to differentiate tilted dipoles whose in-plane angle differs by $180^{\circ}$. \textbf{k} and \textbf{m}, histograms (over 632 and 717 nanospheres, respectively) of the recovered parameters for all the ellipses presented in Figs. \textbf{j} and \textbf{l}.}   
    \label{fig:Fig3}
\end{figure}

\subsection{Experimental results in nonparaxial polarimetry.}
  
C3Pol experiments were performed using as a probe 
an array of 80 nm diameter gold nanospheres, separated by 1.5 $\mu$m to avoid crosstalk between them (Methods). The imaged FOV's diameter is $\sim 36\mu$m.
We first consider test fields resulting from linearly polarized laser beams ($\lambda=532 $ nm) incident just below the critical angle by focusing a
linearly polarized beam at the edge of the objective's back-focal plane (BFP). 
A diaphragm blocks direct reflections 
(Fig. \ref{fig:Fig1}c), so that only the field scattered by the nanospheres
is measured at the detector.  
Figure \ref{fig:Fig3}a shows an image
for a p-polarized incident beam producing a field with
significant axial component (Fig. \ref{fig:Fig3}b). Figure \ref{fig:Fig3}c shows one measured Circ-DSF pair and its fit using the procedure described earlier.  
The retrieved polarizations for all detected Circ-DSFs are nearly rectilinear and mostly axial (Figs. \ref{fig:Fig3}d,e). 
Figures \ref{fig:Fig3}f-i show the corresponding results for an s-polarized incident beam, which produces at the test plane nearly rectilinear polarization with negligible axial component.

The measurement was repeated for eight incident beam directions, as shown in Figs. \ref{fig:Fig3}j and \ref{fig:Fig3}l for incident p- and s-polarization, respectively.  The measured ellipses rotate as expected. The histograms over many nanospheres, displayed in Fig. \ref{fig:Fig3}k for p-polarization, show peak values of $\eta \sim  25 ^{\circ}$, 
$r_{ba} \sim 0.15$, and  
$P \sim 0.9$,
consistent with expected values $(\arccos(0.9)=25.8^\circ$, $r_{ba} = 0$, $P = 1)$.
The histograms for s-polarization (Fig. \ref{fig:Fig3}m) present similar features for $r_{ab}$ and $P$, but $\eta$ peaks at $90^\circ$, as expected.
The widths of these 
histograms are larger than the expected measurement precision, 
probably due to
slight field-dependent polarization aberrations in the setup. 
Nevertheless, these examples show that
C3Pol gives access to  
near-field 3D polarization states at nanoscale locations within the test plane.

\begin{figure}[h]
    \centering
    \includegraphics[scale=0.14]{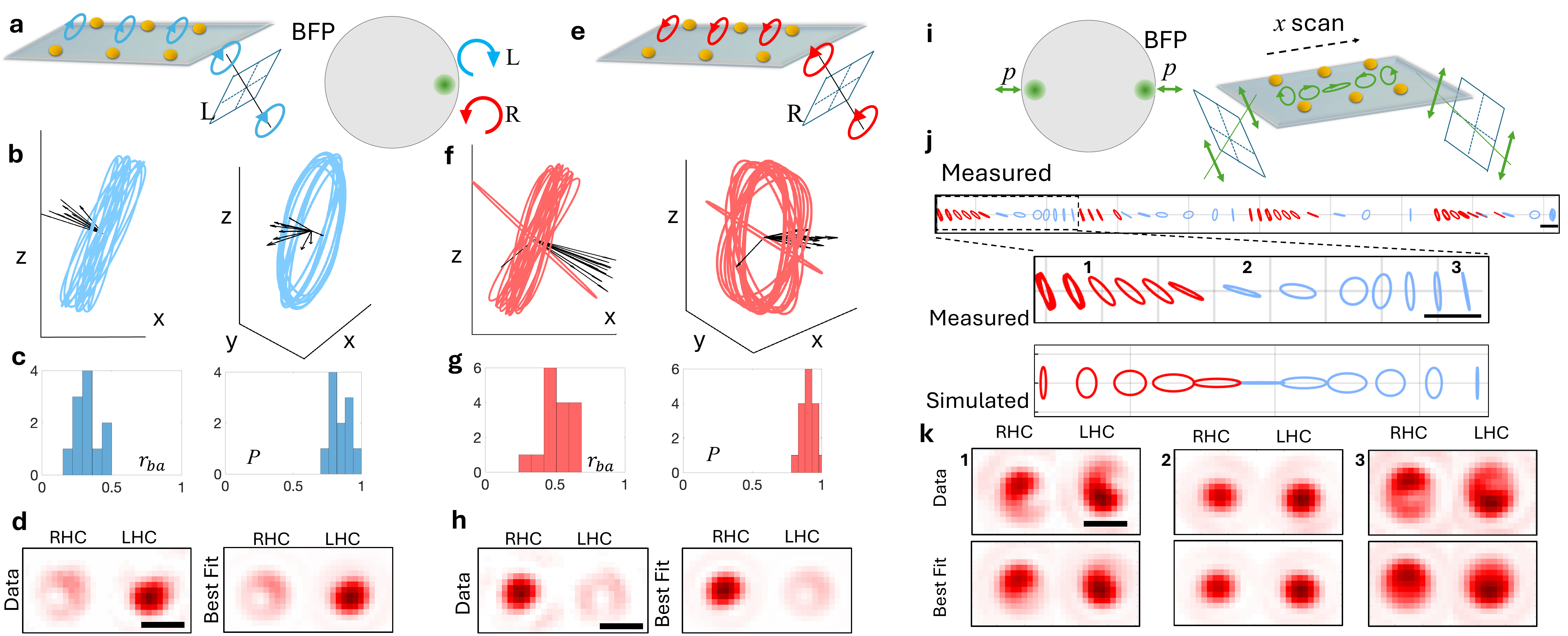}
    \caption{\textbf{C3Pol on elliptically polarized states.} \textbf{a} and \textbf{e}: Illumination schemes using LHC and RHC polarized light (inset: BFP scheme).  The resulting 3D polarizations at the test plane ideally have the same orientation and ellipticity, but opposite spin vectors.
    \textbf{b} and \textbf{f} Superposed recovered polarization ellipses and spin vectors for 11 and 16 nanospheres, respectively, seen from two points of view. 
    \textbf{c} and \textbf{g} Histograms of the measured aspect axis ratio $r_{ba}$ and degree of polarization $P$. 
    \textbf{d} and \textbf{h} Examples of measured Circ-DSFs and their best fit. Scale bar: 365 nm. 
    \textbf{i} Generation of a field with subwavelength polarization variations by focusing two beams at opposite extremes at the BFP of the objective (see BFP scheme). \textbf{j} Measured spatially-varying polarization distributions over the vertical plane.
    The color indicates the sign of $s_{32}$, namely the spin vector's $y$ component. Scale bar 26 nm.  \textbf{k} Examples of measured Circ-DSFs and their corresponding best fit. Scale bar 365 nm. 
    }   
    \label{fig:Fig4}
\end{figure}

To demonstrate C3Pol measurements of 3D spin vectors, we used elliptically polarized incident beams with opposite handedness at the BFP's edge 
(Fig.~\ref{fig:Fig4}a,e).
Figures \ref{fig:Fig4}b,f display 3D views of the retrieved polarization ellipses and their corresponding spin vector (black arrow). As expected, the recovered ellipses 
are similar (Figs. \ref{fig:Fig4}c,g show $r_{ab}$ histograms)  
in both cases, but with opposite handedness so their spin vectors have opposite signs. The retrieved values of $P$ shown in Figs. \ref{fig:Fig4}c,g confirm the high degree of polarization.
Finally, the measured and retrieved Circ-DSFs in Figs. \ref{fig:Fig3}d,h confirm the retrieval's quality.

To generate a field with nanometric-scale polarization variations, two p-polarized beams were applied at opposite edges of the BFP, creating two interfering p-polarized collimated waves producing a periodic polarization pattern at the test plane (Supplementary Note 6). 
The measured period is $\sim 198$ nm, significantly below the particle separation (1.5 $\mu$m). To resolve these subwavelength variations, we captured several images while the particle array is displaced in $x$ with a piezo stage (Methods). The smallest measured separation is $13$ nm. 
Figure \ref{fig:Fig4}j shows over four periods the measured polarization ellipses, whose color indicates the sign of the spin vector's $y$ component. The comparison with the expected field in a zoomed region confirms the transverse spin's sign changes. 
These strong polarization variations are visible from the Circ-DSFs data, some of them shown in Fig. \ref{fig:Fig4}k together with their best fit. 
 
\subsection{C3Pol applied to single-molecule fluorescence microscopy.}

By replacing the nanosphere array with a sample containing fluorescent molecules, C3Pol can be used as a SMOLM technique
to determine the molecules' 3D localization, orientation and degree of wobbling. Since each molecule emits like a rectilinear dipole,  
the three columns of $\bm{\mathcal{I}}$ associated with spin can be ignored, 
reducing possible sources of bias. Due to the excitation absorption probability 
($P_{\rm abs}=\vert \mathbf{E} \cdot \hat{\boldsymbol{\mu}}_{\rm abs} \vert ^2$, $\hat{\boldsymbol{\mu}}_{\rm abs}$ being the absorption dipole), the measured dipoles act as probes of the incoming polarization \cite{novotny2001}. The molecules' direction often fluctuates within the integration time, $P$ is a measure of molecular wobbling (with $P=1$ corresponding to a fixed molecule) \cite{Brasselet:23}. Note that the photon number for single fluorescent molecules is about an order of magnitude below that of the scatterers studied earlier.

\begin{figure}[h]
    \centering
    \includegraphics[scale=0.34]{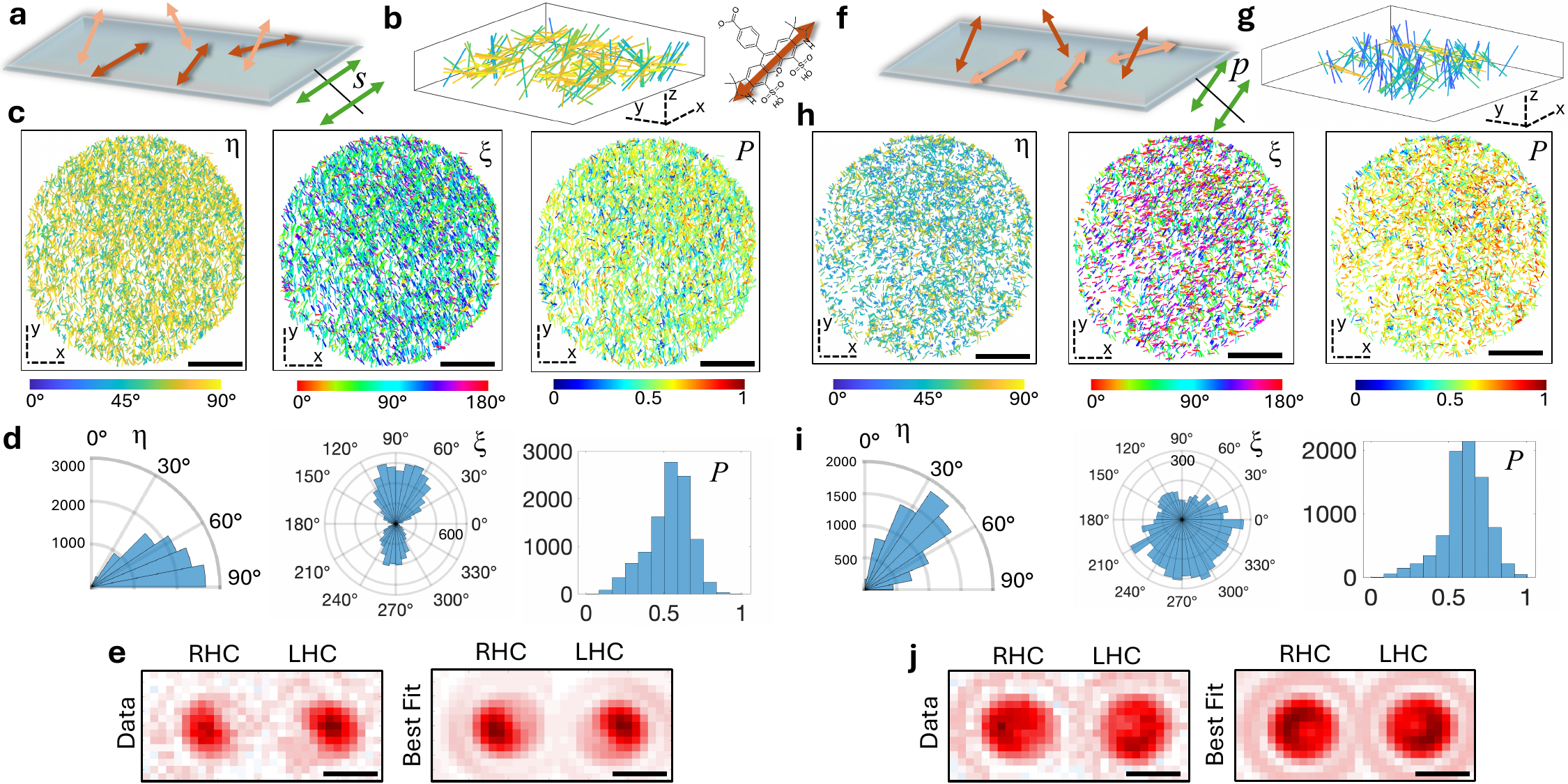}
    \caption{\textbf{Measurement of 3D oriented single fluorophores deposited on a coverslip.} \textbf{a} and \textbf{f} illustration of the s- and p-polarized illuminations. For each of these two illuminations: \textbf{b} and \textbf{g} reconstructed 3D oriented  molecules represented as sticks colored according to the measured polar angle $\eta$. The region of interest is $5.7\times5.7\mu\mathrm{m}^2$. \textbf{c} and \textbf{h} 2D localized molecules across the complete FOV of $\sim 36\mu$m of diameter. The molecules are represented as sticks colored according to the measured angles $\eta$ and $\xi$ (modulo $180^\circ$), and degree of polarization $P$, respectively. The total amount of detected molecules is 10257 and 8138, respectively. Scale bar 9 $\mu$m.  \textbf{d} and \textbf{i} Polar histograms for the recovered angles $\eta$, $\xi$ and histogram for the degree of polarization $P$. \textbf{e} and \textbf{j} Examples of measured Circ-DSFs and their best fit. Scale bar 365 nm. 
    }   
    \label{fig:Fig5}
\end{figure}

We used 
a sparse sample of Alexa Fluor 532 fluorophores immersed in a water solution deposited on a coverslip covered with Polylysine (PLL), and adapted the microscope configuration to fluorescence detection (Methods).   
We first illuminate the sample with an s-polarized laser beam in the TIRF configuration (Fig. \ref{fig:Fig5}a). Figure \ref{fig:Fig5}b shows a 3D view of the reconstructed 3D positions and main 3D directions for the molecules in a small sample region, where each measured fluorophore is represented as a stick whose color indicates the polar angle $\eta$. The detected molecules are oriented close to the sample plane over the whole FOV (Fig. \ref{fig:Fig5}c), as corroborated by the $\eta$ histogram peaking at $60^{\circ}$ (Fig. \ref{fig:Fig5}d). The measured in-plane angles $\xi$ are concentrated around $90^{\circ}$ and $270^{\circ}$ (Fig. \ref{fig:Fig5}c,d), consistent with the excitation polarization, with polar histograms presenting the expected squared cosine distribution. The $P$  
distribution peaks at $\sim 0.5$ (Fig. \ref{fig:Fig5}c,d), 
indicating an amount of wobble consistent with those found with other SMOLM techniques 
\cite{Curcio2020,Nora2023}. 
Figure \ref{fig:Fig5}e shows a 
measured Circ-DSF pair and its best fit, showing C3Pol's robustness even in low-contrast conditions due to noise, background, and depolarization. Figures \ref{fig:Fig5}g-j show similar results for p-polarized illumination, for which the distribution for $P$ behaves similarly but those for the angles change: $\eta$ now peaks at $\sim45^{\circ}$ while $\xi$ is distributed more uniformly, as expected. For each illumination we measured $\sim 10000$ molecules, for which the computational time is only $\sim 50$ seconds (Methods), about an order of magnitude faster than computations based on multiparameter DSF fitting \cite{Brasselet:23}.

\section{Discussion and Conclusion }\label{sec2.1}

We present the implementation of Stokes polarimetry for 3D fields at nanoscales in imaging configuration, extending the widely used Stokes polarimetry techniques to nonparaxial fields. C3Pol gives access to complete nonparaxial polarization information simultaneously at hundreds of positions over micrometer FOVs, reaching nanoscale resolutions 
by scanning the probe. 
Unlike methods based on linear polarization projections \cite{novotny2001,cesarV,Caio:2022,Bruggeman2024}, C3Pol yields unambiguous 3D orientation and spin estimates. It is also appropriate
for SMOLM since it can measure with high precision the 3D orientation and position of single fluorescent molecules. C3Pol can also be extended to other types of dipole scatterers, such as nanorods \cite{nanor}.

Based on the linear dependence of the Circ-DSFs on the SGM  parameters, we implemented a fast polarization retrieval approach in which $\sim 200 $ Circ-DSF pairs can be analyzed per second, 
meeting the large processing needs of SMOLM applications. In contrast, techniques that perform optimization directly on the physical parameters (orientation, ellipticity, wobble) require non-linear optimization \cite{ Ding:2020, Hulleman2021,Zhang2022}. Even approaches based on estimating the matrix elements (linearly related to the DSFs) employ iteration to ensure convergence \cite{Backer_Fourier:2014, Zhang2022}. Similarly, far-field nonparaxial polarization measurements that rely on pupil-plane imaging can only use a single nanoprobe and hence measure one point at a time, unlike C3Pol's imaging configuration that allows measuring polarization at hundreds of locations simultaneously. 

C3Pol is based on a simple setup that uses only standard elements such as wave plates, making it easily implementable with commercial microscopes. Unlike many SMOLM techniques \cite{Brasselet:23}, C3Pol does not require special phase or polarization masks that must be precisely aligned and that can introduce aberrations.
Its essentially diffraction-limited DSFs are also smaller,
allowing higher emitter densities and improving SNR.
Finally by decomposing the Circ-DSFs into SGM basis functions, C3Pol is shown to allow near-optimal 3D polarization retrieval as quantified by the Fisher information. 

\section{Methods}\label{sec3}
\subsection{Experimental setup.}
The setup is based on an inverted microscope (Nikon, Ti-E), using a 100 x magnifying objective (numerical aperture $\mathrm{NA}=1.49$). A laser is used ($\lambda=532$ nm, Oxxius  LCX-532L) with a typical power of 250 mW. For the scattering experiment on gold nanospheres, the beam passes through a secondary setup described in \cite{marco2024} where its amplitude, phase and polarization are modulated, in order to provide a flexible control of the beam at the input pupil plane of the objective. A beam splitter is then used to send the illumination to the test plane: for scattering experiments we use a 50:50 beamsplitter (Thorlabs, BSW4R-532) to transmit the scattered field to the detection path, while for fluorescence experiments we use a dichroic mirror (Semrock, Di02-R532-25x36). At the image plane in the microscope's exit, we place a mechanical iris diaphragm (SF1) to limit the FOV to a diameter of $\sim 36\mu$m. For the telescope we use two 2-inch lenses of focal lengths $f_1=20$cm and $f_2$=25 cm resulting in an extra magnification of $1.25$. The signal is split into its circular polarization components with a QWP (Thorlabs WPQSM05-532 for scattering experiments, achromatic AQWP05M-600 for fluorescence) followed by a Wollaston prism (Thorlabs,WPQ10), resulting in two images that are directly recorded on a CMOS camera (ORCA-Fusion Digital CMOS camera: C14440-20UP) with a typical integration time of 280 ms. The nanosphere array or fluorescent sample are placed on a piezoelectric stage (PI U-781 XY microscope stage). 

For\textit{ scattering experiments}, the reflection of the illumination beams results in spots at the pupil plane that are removed by a second mechanical iris diaphragm (SF2), thereby reducing the effective pupil diameter by a factor of 0.9 (polarization pattern experiment) or 0.85 (linear and elliptical polarizations). To image polarization patterns at high spatial resolution, we measured the scattered signal from each particle at 10 frames/s while they are being displaced along the $x$-direction in steps of $100$ nm. This results in a displacement between images of about 13nm, which is close to the resolution of the method.

For \textit{fluorescence experiments}, the excitation laser is sent to the microscope in a total internal reflexion (TIRF) configuration and the detection numerical aperture is $\mathrm{NA}=1.49$. The illumination polarization is controlled by a half-wave plate. To remove the remaining illumination contributions, we place one emission filter 562/40  under the objective and a second one 561/14 between the camera and the second lens of the telescope.

The quality of the circular polarization detected by the QWP and Wollaston prism is verified by introducing a known circular polarization state in the detection path of the microscope, passing through the 50:50 beam splitter or dichroic mirror (depending on the measurement performed).

\subsection{Nanosphere array probe.}
The array of nanospheres, deposited on a coverslip, was fabricated at the Nano and microfabrication platform PLANETE of the Interdisciplinary Center of Nanoscience of Marseille (CINaM). We use particles of 80 nm of diameter separated by 1.5 $\mu$m. We add a drop of immersion oil (the same that we use for the objective lens) on the nanosphere array and seal it. This way, we create an effective continuous medium between the objective and the nanospheres, so that we can neglect the presence of the glass interface. For this specific particle size and environment, the used laser wavelength of 532 nm corresponds to a high scattering efficiency.

\subsection{Single molecule sample.}
We deposit Poly-L-lysine (PLL, solution, 0.01\%, sterile-filtered, BioReagent) on a plasma-cleaned coverslip. After 15 minutes, we remove the excess of PLL with water and deposit a solution of Alexa Fluor Dye 532 (Thermo Fisher Scientific) in water with a concentration of 1nM. The sample is then sealed.

\subsection{Data analysis.}
\subsubsection{Stokes-Gell-Mann parameter estimation.}

We start the analysis by obtaining a rough estimate of each Circ-DSF's position in the RHC channel. With Matlab's \textit{region props} function we obtain a fast estimation of the DSF's  center, $\mathbf{r}_{\rm RHC}$. We define an image section of $15\times15$ pixels centered at the pixel containing $\mathbf{r}_{\rm RHC}$. Then, we apply an affine transformation $\mathbf{A}$ to $\mathbf{r}_{\rm RHC}$ to find the corresponding position $\mathbf{r}_{\rm LHC}$  in the LHC channel (Supplementary Note 5 and Fig. S15). This affine transformation accounts for translations, rotations, shearing, and scaling between the two coordinate systems. Similarly, we define a section of $15\times15$ pixels centered at the pixel containing $\mathbf{r}_{\rm LHC}$. Finally, we combine the two images into a single $15\times30$ array.  We can write the signal in this combined array as
\begin{equation}
    \mathbf{I}(x,y;S,x_0,y_0,z_0,d)=\alpha B+\sum_{j} S_j \mathcal{I}_{j}(x,y;x_0,y_0,z_0,d),
\end{equation}
where  $(x,y)$ are the array's spatial coordinates, $S_j$ are the SGM parameters, $(x_0,y_0,z_0)$ are the position coordinates of the emitter in the sample volume, $d$ is the interface distance (when there is one) to the focal plane, $B$ is the constant element accounting for the background, and the functions $\mathcal{I}_j$ are the SGM basis elements, which are multiplied by the corresponding SGM parameters $S_j$. While the background $B$ may significantly vary over the full FOV, it is reasonable to assume a constant background over the small regions (of $15 \times 15$ pixels, equivalent to $\approx 790.5\times 790.5  \mathrm{nm}^2$) under analysis. In practice, we set $B(x,y)=\mathrm{max}(\mathcal{I}_0)/8$. This value is chosen such that the \textit{condition number} of the matrix composed solely of the SGM basis functions $\{ \mathcal{I}_j \}$ matches that of the full matrix $\boldsymbol{\mathcal{I}}$.

Let us focus on the estimation of $\boldsymbol{S}$; the Circ-DSFs' dependence on $(x_0,y_0,z_0,d)$ will be discussed in the next section.  We first pile up the columns of $\mathbf{I}$ to form a vector with 450 elements. The same can be done with all basis elements to form a $450 \times 10$ matrix $\boldsymbol{\mathcal{I}}$. The problem is then reduced to the solution of the equation $\mathbf{I}=\boldsymbol{\mathcal{I}} \cdot \boldsymbol{S}$ through the inversion of this rectangular matrix. We numerically verified that the columns of  $\boldsymbol{ \mathcal{I}}$ are linearly independent. The inversion is therefore given by \cite{Boyd2018}  $ \boldsymbol{S}= (\boldsymbol{\mathcal{I}}^{\rm T}\boldsymbol{\mathcal{I}})^{-1} \boldsymbol{\mathcal{I}}^{\rm T}\cdot \mathbf{I}$ where $\boldsymbol{\mathcal{I}}^{\rm T}$ is the transpose of $\boldsymbol{\mathcal{I}}$. The matrix $
(\boldsymbol{\mathcal{I}}^{\rm T}\boldsymbol{\mathcal{I}})^{-1} \boldsymbol{\mathcal{I}}^{\rm T}$ is the Moore–Penrose pseudoinverse.

\subsubsection{Physical constraints applied to the estimation process.}

The estimation given by $(\boldsymbol{\mathcal{I}}^{\rm T}\boldsymbol{\mathcal{I}})^{-1} \boldsymbol{\mathcal{I}}^{\rm T}\cdot \mathbf{I}$  can sometimes lead to estimated vectors $\boldsymbol{S}'$ corresponding to nonphysical polarization states.  For example, these nonphysical results can lead to $\mathbf{DSF}=\sum_jS_j' \mathcal{I}_j$, including some negative intensity values. We exploit the physical constraints of $\mathbf{\Gamma}$ to compute the \textit{closest} physical state to the vector $\boldsymbol{S}'$ . Since  $\mathbf{\Gamma}$ is, by definition, Hermitian and nonnegative-definite, its eigenvalues are nonnegative. We construct the matrix $\mathbf{\Gamma}'$ using equation \ref{eq:Int_S3} and $\boldsymbol{S}'$. We calculate its eigenvalues $\Lambda_1'$, $\Lambda_2'$, $\Lambda_3'$ and the corresponding eigenvectors, $\mathbf{e}_1'$, $\mathbf{e}_2'$, $\mathbf{e}_3'$.
We use the spectral decomposition $\mathbf{\Gamma}'=\mathbf{UDU^{\dagger}}$ where $\mathbf{D}$ is the matrix with the eigenvalues on the diagonal and $\mathbf{U}$ is the matrix containing the eigenvectors as columns.  We then construct the diagonal matrix $\mathbf{D}_{+}$ that contains only the positive diagonal elements of $\mathbf{D}$; the negative ones are set to zero. Hence, the closest physical matrix to $\mathbf{\Gamma}'$ is given by $\mathbf{\mathbf{\Gamma}}=\mathbf{UD_{+}U^{\dagger}}$, as proven in Ref. \cite{realspectral}.

\subsubsection{Estimation of the polarization ellipse.}
To estimate the geometrical aspects of the polarization state characterized by $\boldsymbol{S}$, let us start by writing  $\mathbf{\Gamma} =\sum_j \Lambda_j \mathbf{e}_j\mathbf{e}_j^{\dagger}$, where $\mathbf{e}_j\mathbf{e}_j^{\dagger}$ is the outer product matrix of the eigenvectors  $\mathbf{e}_j$ and $\Lambda_j$ the corresponding eigenvalue. This decomposition is valid for any physical matrix $\mathbf{\Gamma}$. If we assume $\Lambda_1\geq \Lambda_2  \geq \Lambda_3\geq0$, we can  write $\mathbf{\Gamma} $  as \cite{Alonso2023}
\begin{equation}
    \mathbf{\Gamma} =(\Lambda_1-\Lambda_2)\mathbf{e}_1\mathbf{e}_1^{\dagger}+(\Lambda_2-\Lambda_3)\left( \mathbf{e}_1\mathbf{e}_1^{\dagger}+\mathbf{e}_2\mathbf{e}_2^{\dagger}\right)+\Lambda_3 \begin{pmatrix}
1 & 0 & 0 \\
0 & 1 & 0 \\
0 & 0 & 1
\end{pmatrix}.
\label{p:matrix}
\end{equation}
The first term corresponds to a fully polarized state, whereas the last one is associated with a fully unpolarized state. 

If we know that the field is fully polarized, we can set to zero the two smaller eigenvalues, for which the measured values are probably not zero but small (and perhaps negative). This allows us to make the simple association $\bm{\mu}\propto\mathbf{e}_1$. 
We construct then the normalized dipole vector $\mathbf{f}=\mathbf{e}_1e^{-i\Phi}$   where $\Phi=\frac{1}{2}\mathrm{arctan}\left( \mathrm{Im}(\mathbf{e}_1\cdot \mathbf{e}_1)/\mathrm{Re}(\mathbf{e}_1\cdot \mathbf{e}_1)\right)$. The major and minor axis vectors of the ellipse are given by $\mathbf{a}=\mathrm{Re}(\mathbf{f})$ and $\mathbf{b}=\mathrm{Im}(\mathbf{f})$. The spin vector 
is proportional to $2\mathbf{a}\times\mathbf{b}$. 
For the case of a rectilinear dipole, the components of  $\bm{\mu}$ are real and indicate the physical dipole's 3D orientation. We can compute the in-plane  and off-plane angles as $\xi=\mathrm{arctan}(\mu_y/\mu_x)$,  $\eta=\mathrm{arccos}\left(|\mu_z|/\sqrt{\mu_x^2+\mu_y^2+\mu_z^2}\right)$.
For elliptical dipoles, the orientation angles for the major axis are found through the same formulas by simply replacing the components of $\bm{\mu}$ with those of ${\bf a}$. A complete description of the state parametrization is given in the Supplementary Note 2.

If the field is possibly not fully polarized, we simply verify that the retrieved matrix $\mathbf{\Gamma}$ corresponds to a physical state by checking that none of its eigenvalues is negative. If some of the eigenvalues are negative, we set them to zero, as outlined earlier. This leads to a new matrix $\mathbf{\Gamma}$ and a new set of SGM parameters that represent the closest physically realizable polarization state to that resulting from the linear estimator $\bm{\mathcal{I}}^{-1}$. Note that this procedure introduces a slight bias for those states that are near the boundary between physical and unphysical states (Supplementary Fig. S8).

\subsubsection{Localization estimation.}
To account for the dependence on the positions of the dipole and the interface, we generate a library of matrices $\boldsymbol{\mathcal{I}}^{(n)}$, each corresponding to a different set of values for $(x_0,y_0,z_0,d)$ referred to as $(x_0,y_0,z_0,d)^{(n)}$. For the case of a homogeneous medium, $\boldsymbol{\mathcal{I}}^{(n)}$ does not depend on $d$. The calculation of $\boldsymbol{\mathcal{I}}^{(n)}$ accounts for the pupil plane size used for the experiment, including a possible reduction factor in the case of scattering experiments. It also accounts for the used detection wavelengths (532 nm for scattering experiments, 553 nm for fluorescence experiments). To compute each $\boldsymbol{\mathcal{I}}^{(n)}$ we use a pixel size of $52.72$ nm that matches the actual pixel size of the camera. 
In the case of scattering data, we assume $d=0$, whereas for fluorescence $z_0=0$ since we only detect molecules deposited on the coverslip; hence, the basis library only accounts for displacements of the position of the coverslip. Note that in the scattering experiments, the estimation of the spin vector, proportional to $(S_{13},S_{23},S_{33})$, and the 3D position of the particles might be correlated. To experimentally simplify the estimation of the axial position of the nanosphere array, we place it close to the focal plane. 
The library of matrices $\boldsymbol{\mathcal{I}}^{(n)}$ is composed of a set of 625 positions with steps of $(\Delta x_0,\Delta y_0)$ in the $(x, y)$ directions, while five steps are generated in the $z$ direction. The basis sampling for  scattering is $\Delta x_0 =\Delta y_0= 13.2$ nm, $\Delta z_{0}=75$ nm. For fluorescence the sampling is is $\Delta x_0 =\Delta y_0= 26.4$ nm and $\Delta d =75$ nm. 
For each $\boldsymbol{\mathcal{I}}^{(n)}$, we calculate the vector $\boldsymbol{S}^{(n)}$ by multiplying by corresponding pseudo-inverse matrix and generate the $\mathbf{DSF}^{(n)}$, then we quantify the similarity between the measured  $\mathbf{DSF}$ and $\mathbf{DSF}^{(n)}$ with the normalized correlation function $NC= \mathbf{DSF}\cdot \mathbf{DSF}^{(n)}/\left(|\mathbf{DSF}| |\mathbf{DSF}^{(n)}|\right)$. We select the basis $n$ that maximizes $NC$  and associate the corresponding set of coordinates $\boldsymbol{r}_0^{(n)}=(x_0,y_0,z_0,d)^{(n)}$ with the emitter. We use the estimated  $\boldsymbol{S}^{(n)}$ and the $n$-basis to compute the best fit with equation \ref{eq:Eq2}. From Monte Carlo simulations, we observed that by using the basis that best matches the measurements, C3Pol’s estimates maintain the levels of accuracy and precision estimated from centered DSFs.

Finally, the complete data analysis was performed in Matlab with a PC (Intel(R) Xeon(R) W-2235, CPU @ 3.80GHz   3.79 GHz. RAM	32.0 GB. Graphics Card	NVIDIA T400 4GB (4 GB) ) with 6 workers in parallel mode.  \\


\bmhead{Acknowledgements}
The authors thank Guillaume Baffou and the PLANETE platform of CINAM for the access to nanosphere samples, as well as Luis Alem\'an Casta\~neda and Charitra Kumar for fruitful discussions and help on the molecular sample preparation. This research has received funding from the France 2030 investment plan managed by the French National Research Agency (ANR), through the IDEC Equipex+ grant (France 2030 investment plan ANR-21-ESRE-0002), and from the ANR grants 3DPolariSR (ANR-20-CE42-0003) and 3DPol (ANR-21-CE24-0014).\\

\bmhead{Author contributions}
M.A.A. and S.B. conceived the project. I.H., M.A.A. and S.B. conceived the experiment and data analysis. I.H. performed all C3Pol measurements and related data analysis. I.H., M.A.A. and S.B. wrote the manuscript. All
authors contributed to the discussions on the results.\\

\bmhead{Competing interests}
The authors declare no competing interests.\\

\bmhead{Supplementary Information}
Supplementary Notes 1–6, Figs. S1–S15. \\



\begin{thebibliography}{59}


\bibitem[He et~al.(2021)]{He2021}
He, C., He, H., Chang, J., et~al.:
Polarisation optics for biomedical and clinical applications: a review.
\textit{Light: Science \& Applications}. \textbf{10}, 194 (2021).  
\url{https://doi.org/10.1038/s41377-021-00639-x}.

\bibitem[\protect\citeauthoryear{Goldberg}{2022}]{Goldberg2022}
Goldberg, A.Z.:
Chapter Three - Quantum polarimetry.
In: Visser, T.D. (ed.) \textit{Progress in Optics}, vol. 67, pp. 185--274.
Elsevier, Amsterdam (2022).
\url{https://doi.org/10.1016/bs.po.2022.01.001}.

\bibitem[\protect\citeauthoryear{Akiyama et al.}{2021}]{Akiyama_2021}
Akiyama, K., et al.:
First M87 Event Horizon Telescope Results. VII. Polarization of the Ring.
\textit{The Astrophysical Journal Letters}, \textbf{910}(1), L12 (2021).
\url{https://doi.org/10.3847/2041-8213/abe71d}.


\bibitem[\protect\citeauthoryear{Schaefer et al.}{2007}]{Schaefer2007}
Schaefer, B., Collett, E., Smyth, R., Barrett, D., Fraher, B.:
Measuring the Stokes polarization parameters.
\textit{American Journal of Physics}, \textbf{75}(2), 163–168 (2007).
\url{https://doi.org/10.1119/1.2386162}.


\bibitem[\protect\citeauthoryear{Zaidi et~al.}{2024}]{Zaidi2024}
Zaidi, A., Rubin, N.A., Meretska, M.L., et~al.:
Metasurface-enabled single-shot and complete Mueller matrix imaging.
\textit{Nature Photonics}, \textbf{18}, 704--712 (2024).
\url{https://doi.org/10.1038/s41566-024-01426-x}.

\bibitem[\protect\citeauthoryear{Pierangeli and Conti}{2023}]{Pierangeli_2023}
Pierangeli, D., Conti, C., et~al.:
Single-shot polarimetry of vector beams by supervised learning.
\textit{Nature Communications}, \textbf{14}, 1831 (2023).
\url{https://doi.org/10.1038/s41467-023-37474-0}.

\bibitem[\protect\citeauthoryear{Ramkhalawon et~al.}{2013}]{Ramkhalawon:13}
Ramkhalawon, R.D., Brown, T.G., Alonso, M.A., et~al.:
Imaging the polarization of a light field.
\textit{Optics Express}, \textbf{21}(4), 4106–4115 (2013).
\url{https://doi.org/10.1364/OE.21.004106}.

\bibitem[\protect\citeauthoryear{Alonso}{2023}]{Alonso2023}
Alonso, M.A., et~al.:
Geometric descriptions for the polarization of nonparaxial light: a tutorial.
\textit{Advances in Optics and Photonics}, \textbf{15}(1), 176–235 (2023).
\url{https://doi.org/10.1364/AOP.475491}.

\bibitem[\protect\citeauthoryear{Soleillet}{1929}]{Soleillet1929}
Soleillet, P.:
Sur les paramètres caractérisant la polarisation partielle de la lumière dans les phénomènes de fluorescence.
\textit{Annales de Physique}, \textbf{10}(12), 23–97 (1929).
\url{https://doi.org/10.1051/anphys/192910120023}.
\bibitem[\protect\citeauthoryear{Gil et~al.}{2004}]{Gil2004}
Gil, J.J., Correas, J.M., Melero, P.A., Ferreira, C.:
Generalized polarization algebra.
\textit{Monografías del Seminario Matemático García de Galdeano}, \textbf{31}, 161--167 (2004).

\bibitem[\protect\citeauthoryear{Ellis et~al.}{2005}]{Ellis2005}
Ellis, J., Dogariu, A., Ponomarenko, S., et~al.:
Degree of polarization of statistically stationary electromagnetic fields.
\textit{Optics Communications}, \textbf{248}(4-6), 333–337 (2005).
\url{https://doi.org/10.1016/j.optcom.2004.12.050}.


\bibitem[\protect\citeauthoryear{Dennis}{2007}]{Dennis2007}
Dennis, M.R.:
A three-dimensional degree of polarization based on Rayleigh scattering.
\textit{Journal of the Optical Society of America A}, \textbf{24}(7), 2065–2069 (2007).
\url{https://doi.org/10.1364/JOSAA.24.002065}.

\bibitem[\protect\citeauthoryear{Sheppard}{2012}]{Sheppard2012}
Sheppard, C.J.R., et~al.:
Geometric representation for partial polarization in three dimensions.
\textit{Optics Letters}, \textbf{37}(14), 2772–2774 (2012).
\url{https://doi.org/10.1364/OL.37.002772}.


\bibitem[\protect\citeauthoryear{Björk et~al.}{2014}]{Bjrk2014}
Björk, G., de Guise, H., Klimov, A.B., et~al.:
Classical distinguishability as an operational measure of polarization.
\textit{Physical Review A}, \textbf{90}(1), 013830 (2014).
\url{https://doi.org/10.1103/PhysRevA.90.013830}.


\bibitem[\protect\citeauthoryear{Herrera and Quinto-Su}{2023}]{herrera2023measurement}
Herrera, I., Quinto-Su, P.A., et~al.:
Measurement of structured tightly focused beams with classical interferometry.
\textit{Journal of Optics}, \textbf{25}(3), 035602 (2023).
\url{https://doi.org/10.1088/2040-8986/acb44c}.

\bibitem[\protect\citeauthoryear{Quinto-Su}{2023}]{quinto2023interferometric}
Quinto-Su, P.A.:
Interferometric measurement of arbitrary propagating vector beams that are tightly focused.
\textit{Optics Letters}, \textbf{48}(14), 3693–3696 (2023).
\url{https://doi.org/10.1364/OL.492980}.

\bibitem[\protect\citeauthoryear{Kumar et~al.}{2024}]{kumar2024experimental}
Kumar, N., Raju, C., Naik, D.N., Viswanathan, N.K., et~al.:
Experimental measurement of transverse spin dynamics in the non-paraxial focal region.
\textit{Journal of Optics}, \textbf{27}(1), 015608 (2024).
\url{https://doi.org/10.1088/2040-8986/ada047}.


\bibitem[\protect\citeauthoryear{Maluenda et~al.}{2025}]{maluenda2025characterization}
Maluenda, D., Aviñoá, M., Martínez‑Herrero, R., Carnicer, A., et al.:
Characterization of highly focused vector fields using phase retrieval: a practical guide.
\textit{Applied Optics}, \textbf{64}(4), 938–943 (2025).
\url{https://doi.org/10.1364/AO.544562}.


\bibitem[\protect\citeauthoryear{Grosjean et~al.}{2010}]{Grosjean2010}
Grosjean, T., Ibrahim, I.A., Suarez, M.A., Burr, G.W., Mivelle, M., Charraut, D., et~al.:
Full vectorial imaging of electromagnetic light at subwavelength scale.
\textit{Optics Express}, \textbf{18}(6), 5809–5824 (2010).
\url{https://doi.org/10.1364/OE.18.005809}.

\bibitem[\protect\citeauthoryear{Kabakova et~al.}{2016}]{Kabakova2016}
Kabakova, I.V., de Hoogh, A., van der Wel, R.E.C., Wulf, M., le Feber, B., Kuipers, L., et al.:
Imaging of electric and magnetic fields near plasmonic nanowires.
\textit{Scientific Reports}, \textbf{6}, 22665 (2016).
\url{https://doi.org/10.1038/srep22665}.



\bibitem[\protect\citeauthoryear{Neugebauer et~al.}{2018}]{Neugebauer2018}
Neugebauer, M., Eismann, J.S., Bauer, T., Banzer, P.:
Magnetic and Electric Transverse Spin Density of Spatially Confined Light.
\textit{Physical Review X}, \textbf{8}(2), 021042 (2018).
\url{https://doi.org/10.1103/PhysRevX.8.021042}.


\bibitem[\protect\citeauthoryear{le Feber et al.}{2014}]{leFeber2014}
le Feber, B., Rotenberg, N., Beggs, D.M., Kuipers, L., et al.:Simultaneous measurement of nanoscale electric and magnetic optical fields.
\textit{Nature Photonics}, \textbf{8}(1), 43–46 (2014).
\url{https://doi.org/10.1038/nphoton.2013.323}.


\bibitem[\protect\citeauthoryear{le Feber et~al.}{2019}]{leFeber2019}
le Feber, B., Sipe, J.E., Wulf, M., Kuipers, L., Rotenberg, N., et~al.:
A full vectorial mapping of nanophotonic light fields.
\textit{Light: Science \& Applications}, \textbf{8}(1), 28 (2019).
\url{https://doi.org/10.1038/s41377-019-0124-3}.

\bibitem[\protect\citeauthoryear{Rotenberg and Kuipers}{2014}]{Rotenberg2014}
Rotenberg, N., Kuipers, L.:
Mapping nanoscale light fields.
\textit{Nature Photonics}, \textbf{8}, 919--926 (2014).
\url{https://doi.org/10.1038/nphoton.2014.285}.

\bibitem[\protect\citeauthoryear{Frischwasser et~al.}{2021}]{Frischwasser2021}

Frischwasser, K., Cohen, K., Kher‑Alden, J., Dolev, S., Tsesses, S., Bartal, G., et~al.:
Real‑time sub‑wavelength imaging of surface waves with nonlinear near‑field optical microscopy.
\textit{Nature Photonics}, \textbf{15}(6), 442–448 (2021).
\url{https://doi.org/10.1038/s41566-021-00782-2}.
\bibitem[\protect\citeauthoryear{Lee et~al.}{2007}]{Lee2007} 
Lee, K.G., Kihm, H.W., Kihm, J.E., et~al.:
Vector field microscopic imaging of light.
\textit{Nature Photonics}, \textbf{1}, 53--56 (2007).
\url{https://doi.org/10.1038/nphoton.2006.37}.

\bibitem[\protect\citeauthoryear{Ahn et~al.}{2009}]{Ahn:09}
Ahn, J.S., Kihm, H.W., Kihm, J.E., et~al.:
3-Dimensional local field polarization vector mapping of a focused radially polarized beam using gold nanoparticle functionalized tips.
\textit{Optics Express}, \textbf{17}(4), 2280–2286 (2009).
\url{https://doi.org/10.1364/OE.17.002280}.

\bibitem[\protect\citeauthoryear{Bauer et~al.}{2014}]{Bauer:2014}
Bauer, T., Orlov, S., Peschel, U., Banzer, P., Leuchs, G.:
Nanointerferometric amplitude and phase reconstruction of tightly focused vector beams.
\textit{Nature Photonics}, \textbf{8}, 23--27 (2014).
\url{https://doi.org/10.1038/nphoton.2013.289}.

\bibitem[\protect\citeauthoryear{Bauer et~al.}{2015}]{Bauer:2015}

Bauer, T., Banzer, P., Karimi, E., Orlov, S., Rubano, A., Marrucci, L., Santamato, E., Boyd, R.W., Leuchs, G.:
Observation of optical polarization Möbius strips.
\textit{Science}, \textbf{347}(6225), 964–966 (2015).
\url{https://doi.org/10.1126/science.1260635}.

\bibitem[\protect\citeauthoryear{Yang et~al.}{2023}]{Yang2023}
Yang, D., Hu, H., Gao, H., Chen, J., Zhan, Q.:
Mie scattering nanointerferometry for the reconstruction of tightly focused vector fields by polarization decomposition.
\textit{Photonics}, \textbf{10}(5), 0496 (2023).
\url{https://doi.org/10.3390/photonics10050496}.

\bibitem[\protect\citeauthoryear{Marchenko et~al.}{2011}]{Marchenko2011}

Marchenko, P., Orlov, S., Huber, C., Banzer, P., Quabis, S., Peschel, U., Leuchs, G.:
Interaction of highly focused vector beams with a metal knife-edge.
\textit{Optics Express}, \textbf{19}(8), 7244–7261 (2011).
\url{https://doi.org/10.1364/OE.19.007244}.

\bibitem[\protect\citeauthoryear{Huber et~al.}{2013}]{Huber2013}
Huber, C., Orlov, S., Banzer, P., Leuchs, G.:
Corrections to the knife-edge based reconstruction scheme of tightly focused light beams.
\textit{Optics Express}, \textbf{21}(21), 25069–25076 (2013).
\url{https://doi.org/10.1364/OE.21.025069}.

\bibitem[\protect\citeauthoryear{Grosjean and Courjon}{2006}]{Grosjean:06}
Grosjean, T., Courjon, D.:
Photopolymers as vectorial sensors of the electric field.
\textit{Optics Express}, \textbf{14}(6), 2203–2210 (2006).
\url{https://doi.org/10.1364/OE.14.002203}.

\bibitem[\protect\citeauthoryear{Porfirev et~al.}{2022}]{Porfirev2022}

Porfirev, A., Khonina, S., Ivliev, N., Meshalkin, A., Achimova, E., Forbes, A.:
Writing and reading with the longitudinal component of light using carbazole-containing azopolymer thin films.
\textit{Scientific Reports}, \textbf{12}, 3477 (2022).
\url{https://doi.org/10.1038/s41598-022-07440-9}.

\bibitem[\protect\citeauthoryear{Otte et~al.}{2019}]{Otte:2019}

Otte, E., Tekce, K., Lamping, S., Ravoo, B.J., Denz, C.:
Polarization nano-tomography of tightly focused light landscapes by self-assembled monolayers.
\textit{Nature Communications}, \textbf{10}(1), 4308 (2019).
\url{https://doi.org/10.1038/s41467-019-12127-3}.

\bibitem[\protect\citeauthoryear{Sick et~al.}{2001}]{Sick2000}
Sick, B., Hecht, B., Wild, U.P., Novotny, L.:
Probing confined fields with single molecules and vice versa.
\textit{Journal of Microscopy}, \textbf{202}(Pt 2), 365–373 (2001).
\url{https://doi.org/10.1046/j.1365-2818.2001.00795.x}.

\bibitem[\protect\citeauthoryear{Novotny et~al.}{2001}]{novotny2001}

Novotny, L., Beversluis, M.R., Youngworth, K.S., Brown, T.G.:
Longitudinal field modes probed by single molecules.
\textit{Physical Review Letters}, \textbf{86}, 5251–5254 (2001).
\url{https://doi.org/10.1103/PhysRevLett.86.5251}.

\bibitem[\protect\citeauthoryear{Lieb et~al.}{2004}]{Lieb2004}
Lieb, M.A., Zavislan, J.M., Novotny, L.:
Single‑molecule orientations determined by direct emission pattern imaging.
\textit{Journal of the Optical Society of America B}, \textbf{21}(6), 1210–1215 (2004).
\url{https://doi.org/10.1364/JOSAB.21.001210}.

\bibitem[\protect\citeauthoryear{Yu et~al.}{2022}]{Yu2022}

Yu, T., Rodriguez, F., Schedin, F., Kravets, V.G., Zenin, V.A., Bozhevolnyi, S.I., Novoselov, K.S., Grigorenko, A.N.:
Nanoscale light field imaging with graphene.
\textit{Communications Materials}, \textbf{3}(1), 40 (2022).
\url{https://doi.org/10.1038/s43246-022-00264-0}.




\bibitem[\protect\citeauthoryear{Eismann et~al.}{2020}]{Eismann2020}

Eismann, J.S., Nicholls, L.H., Roth, D.J., Alonso, M.A., Banzer, P., Rodríguez‑Fortuño, F.J., Zayats, A.V., Nori, F., Bliokh, K.Y.:
Transverse spinning of unpolarized light.
\textit{Nature Photonics}, \textbf{15}(2), 156–161 (2020).
\url{https://doi.org/10.1038/s41566-020-00733-3}.

\bibitem[\protect\citeauthoryear{Eismann and Banzer}{2025}]{elec_mag}
Eismann, J.S., Banzer, P.:
Nanoscale vectorial electric and magnetic field measurement.
\textit{ACS Photonics}, \textbf{12}(1), 522–527 (2025).
\url{https://doi.org/10.1021/acsphotonics.4c01831}.

\bibitem[\protect\citeauthoryear{Brasselet and Alonso}{2023}]{Brasselet:23}
Brasselet, S., Alonso, M.A.:
Polarization microscopy: from ensemble structural imaging to single‑molecule 3D orientation and localization microscopy.
\textit{Optica}, \textbf{10}(11), 1486–1510 (2023).
\url{https://doi.org/10.1364/OPTICA.502119}.

\bibitem[\protect\citeauthoryear{Herrera et~al.}{2024}]{Herrera2024}
Herrera, I., Alemán‑Castañeda, L.A., Brasselet, S., Alonso, M.A.:
Stokes‑based analysis for the estimation of 3D dipolar emission.
\textit{Journal of the Optical Society of America A}, \textbf{41}(11), 2134–2148 (2024).
\url{https://doi.org/10.1364/JOSAA.538706}.

\bibitem[\protect\citeauthoryear{Curcio et~al.}{2020}]{Curcio2020}
Curcio, V., Alemán‑Castañeda, L.A., Brown, T.G., Brasselet, S., Alonso, M.A., et~al.:
Birefringent Fourier filtering for single molecule coordinate and height super‑resolution imaging with dithering and orientation.
\textit{Nature Communications}, \textbf{11}(1), 2279 (2020).
\url{https://doi.org/10.1038/s41467-020-19064-6}.

\bibitem[\protect\citeauthoryear{Goudail}{2017}]{Goudail2017}
Goudail, F.:
Performance comparison of pseudo‑inverse and maximum‑likelihood estimators of Stokes parameters in the presence of Poisson noise for spherical design‑based measurement structures.
\textit{Optics Letters}, \textbf{42}(10), 1899 (2017).
\url{https://doi.org/10.1364/ol.42.001899}.

\bibitem[\protect\citeauthoryear{Vio et~al.}{2005}]{Vio2005}

Vio, R., Bardsley, J., Wamsteker, W.:
Least‑squares methods with Poissonian noise: Analysis and comparison with the Richardson‑Lucy algorithm.
\textit{Astronomy \& Astrophysics}, \textbf{436}(2), 741–755 (2005).
\url{https://doi.org/10.1051/0004-6361:20041997}.

\bibitem[\protect\citeauthoryear{Munger et~al.}{2023}]{Nora2023}

Munger, E., Sison, M., Brasselet, S.:
Influence of the excitation polarization on single molecule 3D orientation imaging.
\textit{Optics Communications}, \textbf{541}, 129480 (2023).
\url{https://doi.org/10.1016/j.optcom.2023.129480}

\bibitem[\protect\citeauthoryear{Cruz et~al.}{2016}]{cesarV}
Cruz, C.A.V., Shaban, H.A., Kress, A., Bertaux, N., Monneret, S., Mavrakis, M., Savatier, J., Brasselet, S.:
Quantitative nanoscale imaging of orientational order in biological filaments by polarized super‑resolution microscopy.
\textit{Proceedings of the National Academy of Sciences}, \textbf{113}(7), 820–828 (2016).
\url{https://doi.org/10.1073/pnas.1516811113}.

\bibitem[\protect\citeauthoryear{Rimoli et~al.}{2022}]{Caio:2022}
Rimoli, C.V., Valades-Cruz, C.A., Curcio, V., Mavrakis, M., Brasselet, S., et~al.:
4polar‑STORM polarized super‑resolution imaging of actin filament organization in cells.
\textit{Nature Communications}, \textbf{13}(1), 301 (2022).
\url{https://doi.org/10.1038/s41467-022-27966-w}.

\bibitem[\protect\citeauthoryear{Bruggeman et~al.}{2024}]{Bruggeman2024}

Bruggeman, E., Zhang, O., Needham, L.-M., Körbel, M., Daly, S., Cheetham, M., Peters, R., Wu, T., Klymchenko, A.S., Davis, S.J., Paluch, E.K., Klenerman, D., Lew, M.D., O’Holleran, K., Lee, S.F.:
PolCam: instant molecular orientation microscopy for the life sciences.
\textit{Nature Methods}, \textbf{21}(10), 1873–1883 (2024).
\url{https://doi.org/10.1038/s41592-024-02382-8}.


\bibitem[\protect\citeauthoryear{Bouchal et~al.}{2023}]{nanor}
Bouchal, P., Dvořák, P., Hrtoň, M., Rovenská, K., Chmelík, R., Šikola, T., Bouchal, Z.:
Single‑shot aspect ratio and orientation imaging of nanoparticles.
\textit{ACS Photonics}, \textbf{10}(9), 3331–3341 (2023).
\url{https://doi.org/10.1021/acsphotonics.3c00785}.

\bibitem[\protect\citeauthoryear{Ding et~al.}{2020}]{Ding:2020}
Ding, T., Wu, T., Mazidi, H., Zhang, O., Lew, M.D.:
Single‑molecule orientation localisation microscopy for resolving structural heterogeneities between amyloid fibrils.
\textit{Optica}, \textbf{7}(6), 602–607 (2020).
\url{https://doi.org/10.1364/OPTICA.388157}.
\bibitem[\protect\citeauthoryear{Hulleman et~al.}{2021}]{Hulleman2021}
Hulleman, C.N., Thorsen, R.Ø., Kim, E., et~al.:
Simultaneous orientation and 3D localization microscopy with a vortex point spread function.
\textit{Nature Communications}, \textbf{12}(1), 5934 (2021).
\url{https://doi.org/10.1038/s41467-021-26228-5}.

\bibitem[\protect\citeauthoryear{Zhang et~al.}{2022}]{Zhang2022}

Zhang, O., Guo, Z., He, Y., Wu, T., Vahey, M.D., Lew, M.D.:
Six‑dimensional single‑molecule imaging with isotropic resolution using a multi‑view reflector microscope.
\textit{Nature Photonics}, \textbf{17}(2), 179–186 (2022).
\url{https://doi.org/10.1038/s41566-022-01116-6}.

\bibitem[\protect\citeauthoryear{Backer and Moerner}{2014}]{Backer_Fourier:2014}
Backer, A.S., Moerner, W.E., et~al.:
Extending single‑molecule microscopy using optical Fourier processing.
\textit{The Journal of Physical Chemistry B}, \textbf{118}(28), 8313–8329 (2014).
\url{https://doi.org/10.1021/jp501778z}.

\bibitem[\protect\citeauthoryear{Marco et~al.}{2024}]{marco2024}

Marco, D., Herrera, I., Brasselet, S., Alonso, M.A.:
Periodic skyrmionic textures via conformal cartographic projections.
\textit{APL Photonics}, \textbf{9}(11), 110803 (2024).
\url{https://doi.org/10.1063/5.0230959}.

\bibitem[\protect\citeauthoryear{Boyd and Vandenberghe}{2018}]{Boyd2018}

Boyd, S., Vandenberghe, L.:
Introduction to Applied Linear Algebra: Vectors, Matrices, and Least Squares.
\textit{Cambridge University Press}, (2018).
\url{http://dx.doi.org/10.1017/9781108583664}.

\bibitem[\protect\citeauthoryear{Cheng and Higham}{1998}]{realspectral}
Cheng, S.H., Higham, N.J.:
A modified Cholesky algorithm based on a symmetric indefinite factorization.
\textit{SIAM Journal on Matrix Analysis and Applications}, \textbf{19}(4), 1097–1110 (1998).
\url{https://doi.org/10.1137/S0895479896302898}.

\end{thebibliography}
\end{document}